\documentclass[12pt,preprint]{aastex}

\shorttitle{Tuning Gravitationally Lensed Standard Sirens}
\shortauthors{J\"onsson et al.}

\begin{document}

\title{TUNING GRAVITATIONALLY LENSED STANDARD SIRENS}

\author{
  J.~J\"onsson and
  A.~Goobar}
\affil{Stockholm University, AlbaNova University Center, \\
  Fysikum, SE-10691 Stockholm, Sweden}
\and
\author{E.~M\"ortsell}
\affil{Stockholm University, AlbaNova University Center, \\
  Stockholm Observatory, SE-10691 Stockholm, Sweden}
\email{jacke@physto.se}

\begin{abstract}
Gravitational waves emitted by chirping supermassive black hole
binaries could in principle be used to obtain very accurate distance
determinations. Provided they have an electromagnetic counterpart from
which the redshift can be determined, these standard sirens could be
used to build a high redshift Hubble diagram. Errors in the distance
measurements will most likely be dominated by gravitational lensing. 
We show that the (de)magnification due to inhomogeneous foreground matter 
will increase the scatter in the measured distances by a factor $\sim 10$.
We propose to use optical and IR data of the foreground galaxies to 
minimize the degradation from weak lensing. 
We find that the net effect of correcting
the estimated distances for lensing is comparable to
increasing the sample size by a factor of three when using the data to
constrain cosmological parameters.
\end{abstract}


\section{INTRODUCTION}
The distance--redshift relation for electromagnetic standard candles,
such as Type Ia supernovae, has been crucial to our understanding of
the energy content of the Universe. With the advent of laser
interferometric gravitational wave detectors, such as
LIGO\footnote{\tt http://www.ligo.caltech.edu/}
and the planned space mission LISA\footnote{\tt
http://lisa.jpl.nasa.gov/}, new windows to the Universe will be
opened. Binary systems with shrinking orbits, emitting gravitational
waves just before coalescence, are expected to be standard sirens, the
gravitational wave analogy of electromagnetic standard candles
\citep{sch86,sch02}. Since the frequency of the gravitational waves
rises when the orbit changes, these systems are also known as chirping
binaries \citep{sch86}. Laser interferometers can provide crude
estimates of the position of the gravitational wave source in the
sky. If a source has an observable optical counterpart, the position
of the siren could be pinpointed and the distance measurement
significantly improved
\citep{sch86}.

Accurate redshifts of electromagnetic standard candles are usually
easier to obtain than accurate distance measurements. For
gravitational wave standard sirens the situation is reversed: accurate
distances, but no redshifts, can be obtained. An electromagnetic
counterpart could thus not only increase the accuracy in the distance
measurable by LISA to $<1\%$ \citep{hol05} but also provides the
redshift. Detection of gravitational wave sources with electromagnetic
counterparts thus have the potential to probe the distance--redshift
relation with unprecedented precision 
\citep{sch86}. For a recent discussion on possible
observable electromagnetic counterparts of massive black hole
binaries, see \citet{dot06} and references therein.

The coalescence of supermassive black holes is associated with mergers
of proto-galaxies or galaxies, many of which are expected to occur at high
redshift. Chirping supermassive binary black holes thus promise
to be probes of the very distant Universe. 
The exact redshift distribution of supermassive binary black
holes is, however, poorly known and the estimated number of events
detectable by LISA range from a few to hundreds of thousands per year
\citep{men01,hae03,Wyi03,eno04,isl04,ses04,rho05,ses05,kou06}. 
While cosmological parameters can be derived from a Hubble diagram of
``standard sirens'' at high $z$, severe parameter degeneracies result
unless there are also data points at low redshifts. Since the
distances which could be obtained from chirping supermassive black
hole binaries are absolute, they could be combined with absolute
distances measured using other low redshift probes, e.g.~chirping
neutron star binaries \citep{dal06}.
At high redshifts, distance measurements will be affected by
gravitational deflection of the radiation by foreground matter
\citep{mar93}. An additional scatter of a factor $\sim 10$ is
thus expected due to gravitational lensing (see Section \ref{sec:flux}).

In principle, due to flux conservation, the effects of gravitational
lensing are expected to average out if a large number of standard
candles or sirens are observed. However, the large uncertainty in the
rate, coupled with the fact that it might be difficult to find and
observe electromagnetic counterparts, suggests that a scenario where
only a handful of standard sirens with redshifts can be measured by
LISA is not unlikely. The scatter due to gravitational lensing will
then limit the usefulness of the data severely.

In this paper, we propose a method to compensate for the degradation
due to lensing, at least partially.
\citet{gun06} showed that properties of foreground galaxies can be used
to estimate the magnification accurately enough for corrections of
individual sources to be useful. The error box in the sky provided by
LISA, will probably be extensively observed in many different
wavelength bands in the search for an electromagnetic
counterpart. This data could be used to estimate the magnification of
the siren due to gravitational lensing by foreground matter. For
reasonable detection rates of chirping supermassive binary black
holes, we show that correcting for gravitational lensing will be
comparable to increasing the sample size by a factor three. The
corrections will thus be crucial when using the data to constrain
cosmological parameters such as the Hubble constant, the matter
density and dark energy properties.
The cosmology used throughout this paper is the concordance 
flat $\Lambda$CDM model, $[h,\Omega_{\rm
M},w_0]=[0.7,0.3,-1]$.

\section{CORRECTING GRAVITATIONALLY LENSED SOURCES}
In this section we describe how standard sirens can be corrected for
gravitational lensing.

\subsection{Computing Corrections}
As noted, gravitational waves emitted from high redshift standard
sirens might be gravitationally deflected by matter along the
line of sight. The magnification factor, $\mu$, describes the 
amplification ($\mu>1$) or
de-amplification ($\mu<1$) of a source due to gravitational lensing as compared
to a universe with homogeneously distributed matter. If $D_{\rm L}$ is
the luminosity distance in a homogeneous universe (which is what is
used when estimating cosmological parameters), the observed luminosity
distance, $D_{\rm L}^{\rm obs}$, affected by gravitational lensing, is
$\mu^{-1/2}D_{\rm L}$. 

\citet{gun06} showed that observed properties 
(e.g.~position, spectroscopic or photometric redshift, and luminosity) of
intervening galaxies along the line-of-sight of the source can be used
to estimate the magnification factor with sufficient accuracy for
corrections to be useful. 
The magnification factor can be computed using halo profiles depending
on the mass of the galaxy halo, which can be obtained from the galaxy
luminosity using Tully--Fisher and Faber--Jackson relations.
This method has been successfully applied to,
e.g.~33 supernovae in the GOODS fields \citep{jon06}.  The
luminosity distance corrected for gravitational lensing using the
estimated magnification factor, $\mu_{\rm est}$, is 
\begin{equation}
D_{\rm L}^{\rm corr}=\mu_{\rm est}^{1/2}D_{\rm L}^{\rm
obs}=(\mu/\mu_{\rm est})^{-1/2}D_{\rm L}.
\end{equation} 
Whether the corrected distance is closer to the $D_{\rm L}$ obtained
in a homogeneous universe than the observed distance depends on how
close the quantity $\mu/\mu_{\rm est}$ is to unity, i.e.~how well we
can estimate the magnifications. This in turn depends on the
completeness and accuracy of the foreground galaxy information.
We have used the method presented in \citet{gun06} to
obtain the precision to which $\mu_{\rm est}$ can be estimated for
sources out to $z=4$. 
The distribution of $\mu/\mu_{\rm est}$ at a specific redshift was
found by computing $\mu$ and $\mu_{\rm est}$ for a large number of
simulated lines-of-sight.  
Galaxies along each line-of-sight were simulated and the magnification
computed using a fiducial model. This value was taken as the 
true magnification $\mu$. 
Since the mean of a simulated probability distribution function
(PDF) of $\mu$ is not exactly unity, each PDF of $\mu$ was normalized to
ensure a mean of unity.
A value of the magnification
representing $\mu_{\rm est}$ was also computed assuming a limiting
galaxy magnitude and a reasonable
error budget including the effects of using the wrong lens model 
(for 50\% of the lenses a SIS-profile was used instead of a NFW-profile)
and uncertainties in redshifts and positions as well as intrinsic 
scatter and accuracy in the Tully--Fisher and Faber--Jackson relations.
Since new powerful instruments for
electromagnetic observations should be available when LISA can be
expected to be operational, our simulations assume galaxy optical
limiting magnitude of $I=28$.
In these simulations we have only considered the gravitational
deflection due to galaxies and neglected any effects of large scale 
structures.

\subsection{Distance Probability Distribution Functions}\label{sec:dpdf}
The accuracy to which the luminosity distance to chirping supermassive
binary black holes could be measured by LISA, neglecting the effects
from lensing, has been investigated by \citet{hol05}. In
their most optimistic scenario, where an electromagnetic counterpart
was assumed, the fractional error in the distance measured for sirens
at redshift $z=1$ and $z=3$ was found to be $\delta D_{\rm L}/D_{\rm
L} \sim 0.1\%$ and $\delta D_{\rm L}/D_{\rm L} \sim 0.5\%$,
respectively. We use these numbers and linear interpolation and
extrapolation to obtain the fractional errors used in subsequent
calculations. The PDF describing
the dispersion in the distance obtained from an 
unlensed standard siren is assumed to be a Gaussian.

Supermassive binary black holes thus have the potential to be almost
perfect (absolute) standard sirens and the gain from being able to
minimize any additional scatter, e.g. lensing, can be very large.

The PDF of the observed luminosity
distance $D_{\rm L}^{\rm obs}=\mu^{-1/2} D_{\rm L}$ is given by the
convolution of the PDFs of $\mu^{-1/2}$ and $D_{\rm L}$. Both PDFs
depend on redshift and, to some extent, on 
cosmological parameters. Corrected luminosity
distances are expected to be distributed according to the PDF given by
the convolution of the PDFs of $(\mu/\mu_{\rm est})^{-1/2}$ and
$D_{\rm L}$. 
The cosmology dependence
on the magnification factor is weak and is therefore neglected.

\begin{figure}
\begin{center}
\resizebox{0.45\textwidth}{!}{\includegraphics{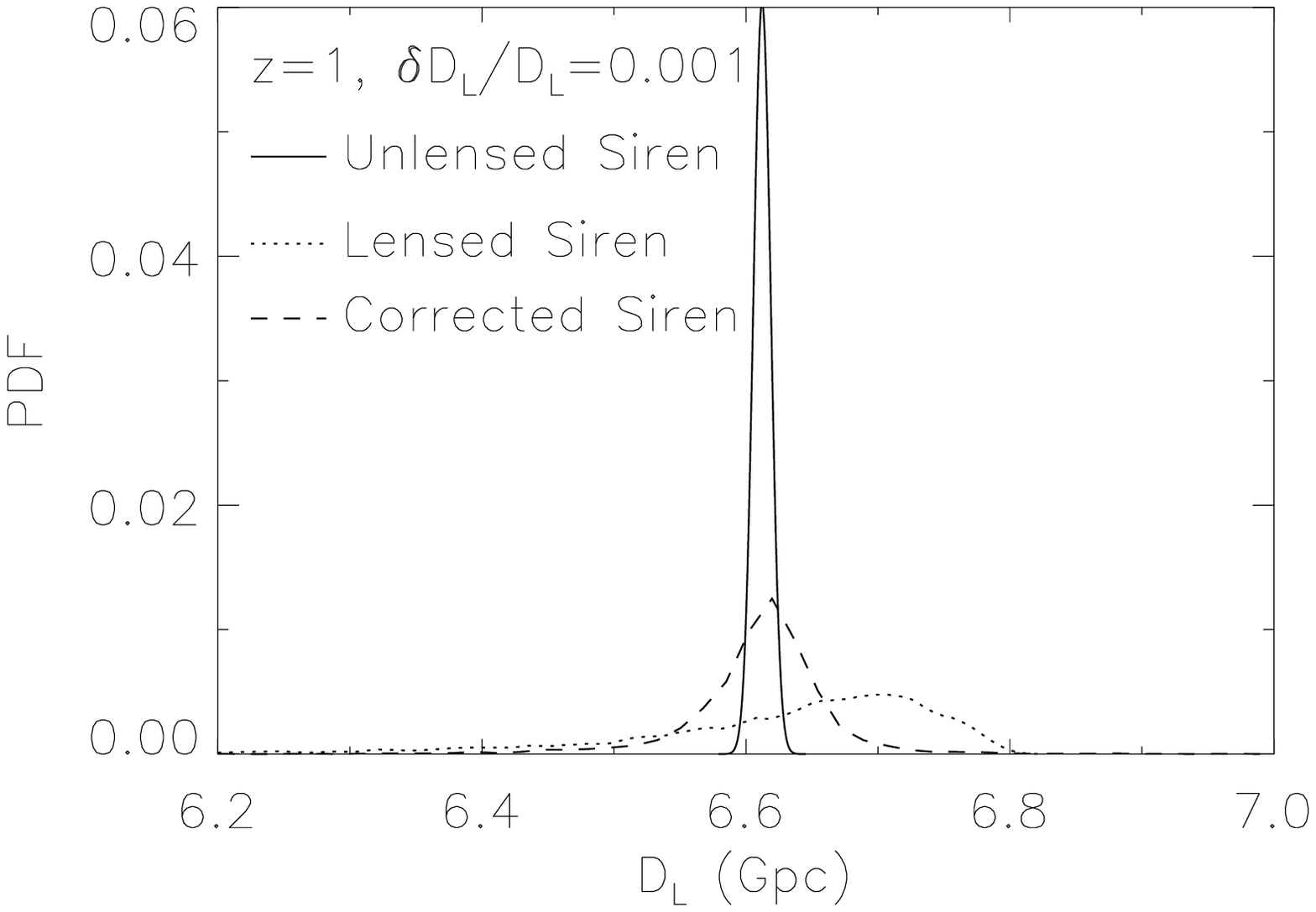}}
\resizebox{0.45\textwidth}{!}{\includegraphics{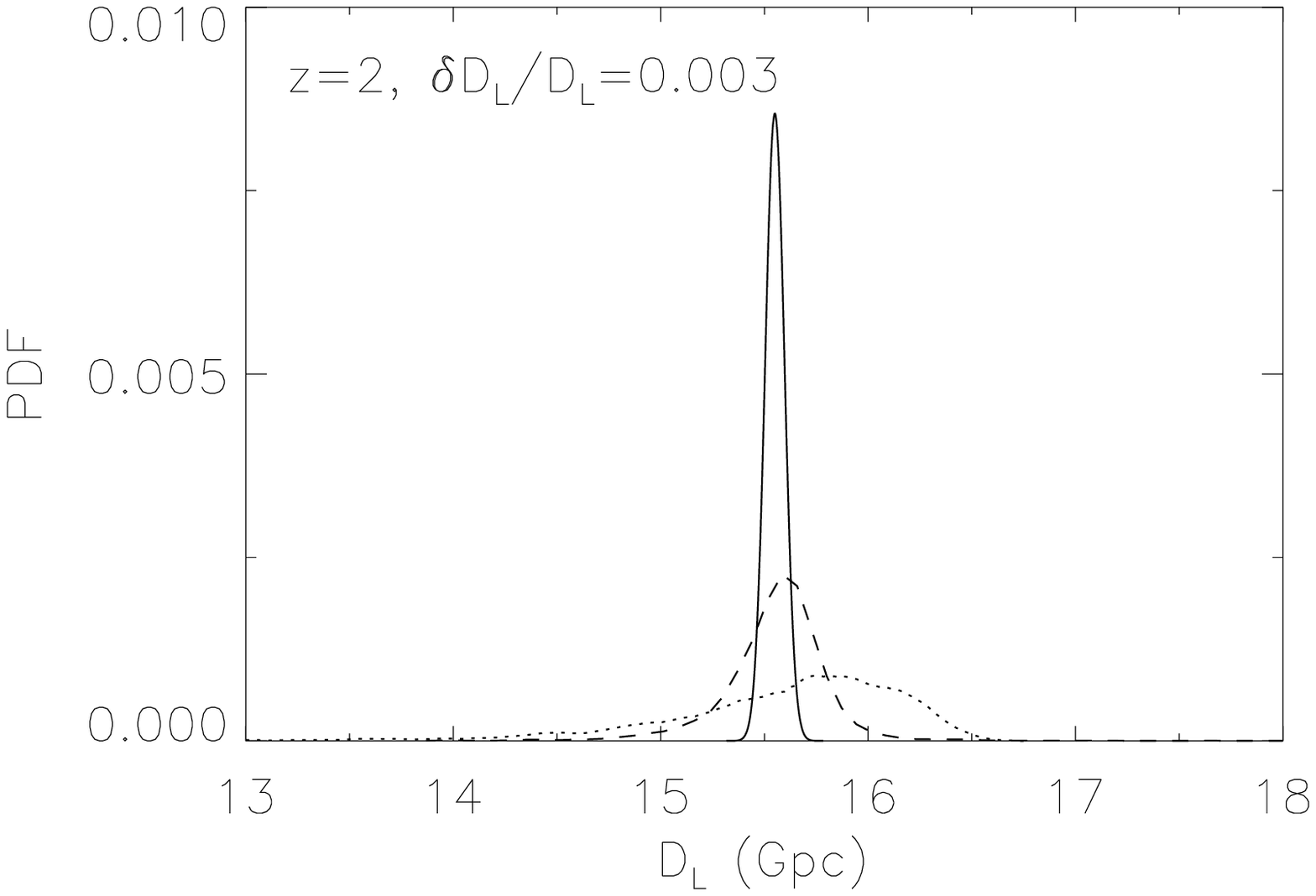}}
\resizebox{0.45\textwidth}{!}{\includegraphics{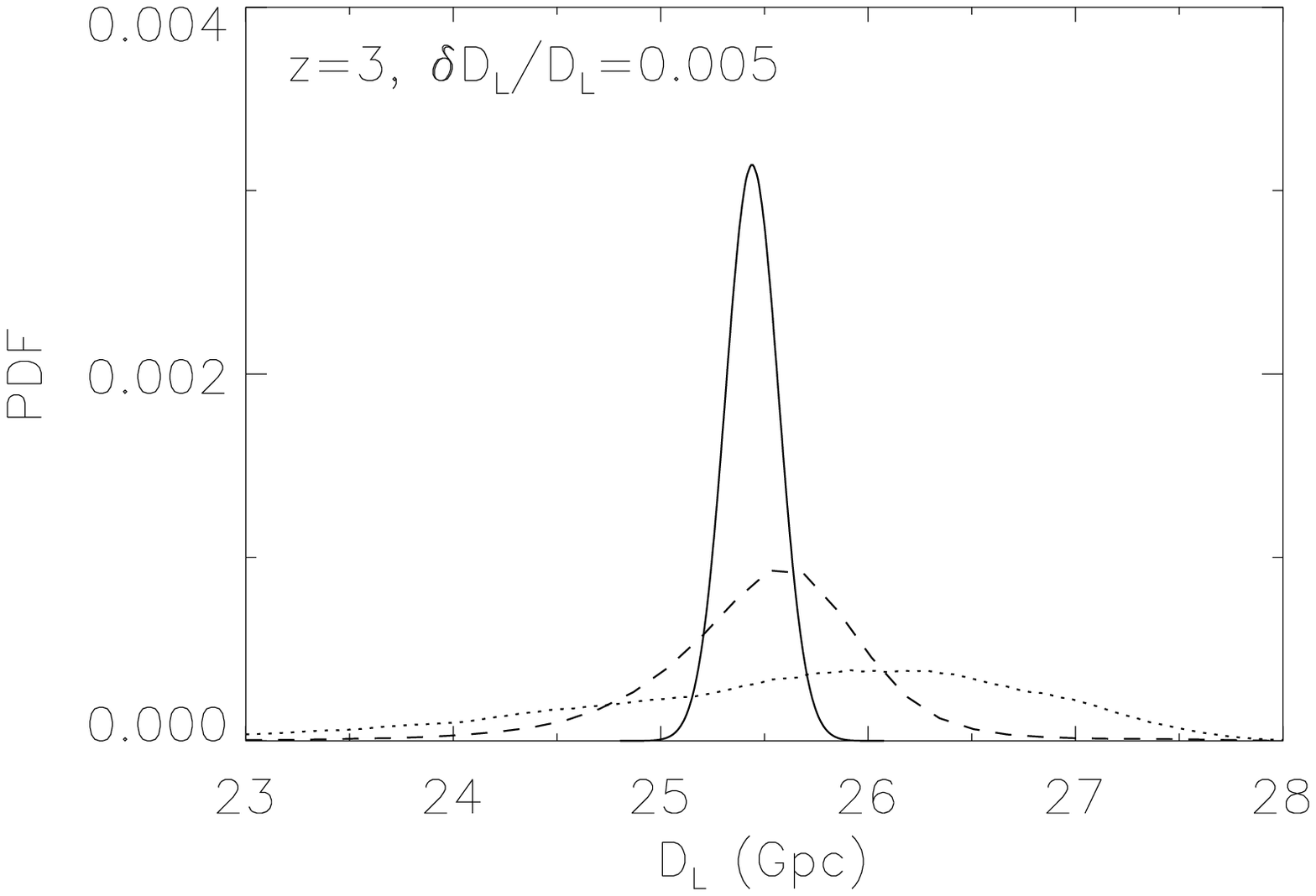}}
\resizebox{0.45\textwidth}{!}{\includegraphics{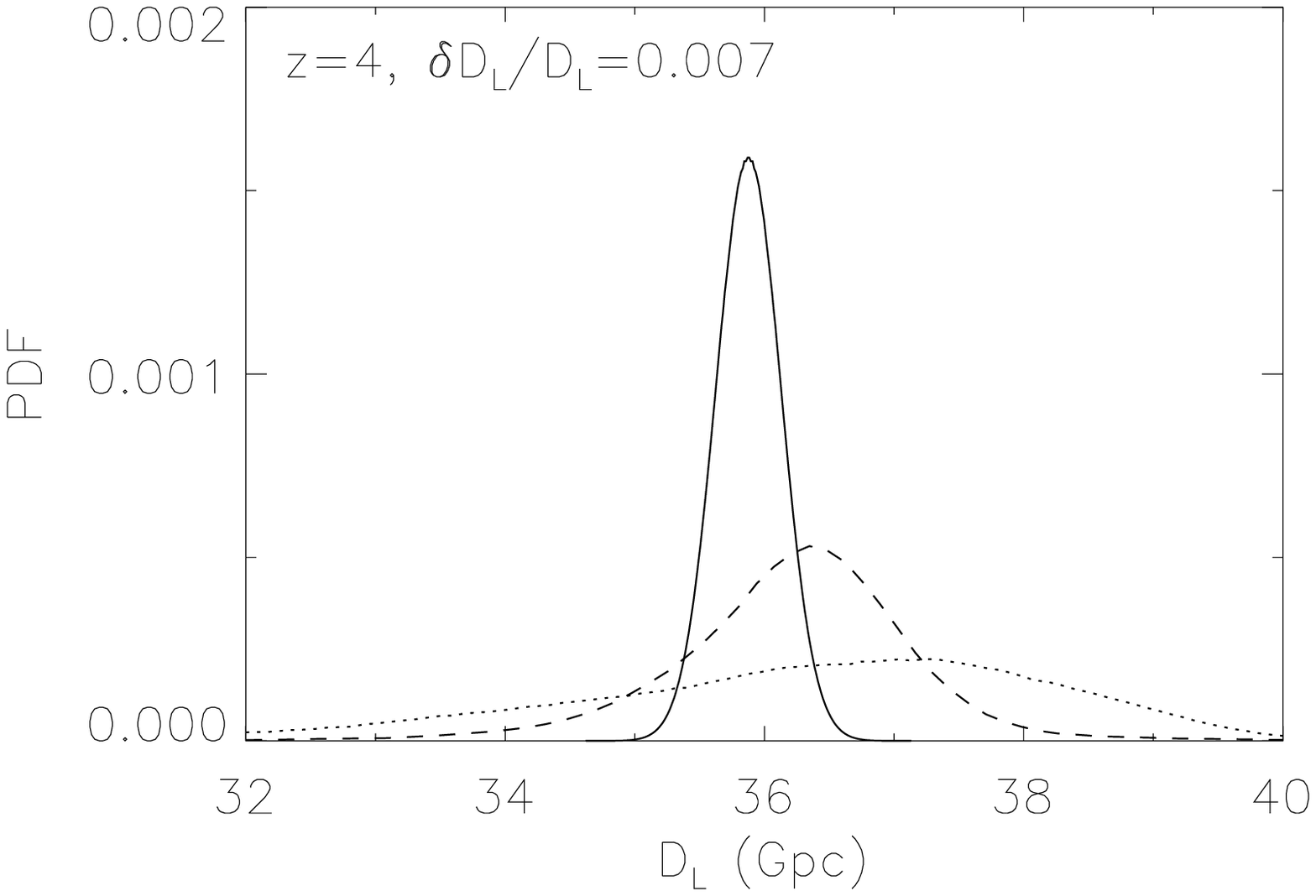}}
\caption{\label{fig:pdf} Probability distribution functions (PDF) for
  luminosity distances obtainable from unlensed, lensed, and corrected
  standard sirens at different redshifts. The fractional uncertainty to which
  the luminosity distance to an unlensed siren could be measured by
  LISA is indicated in each panel.}
\end{center}
\end{figure}

Figure \ref{fig:pdf} shows the PDFs of luminosity distances measured
from unlensed, lensed, and corrected standard sirens at different
redshifts.  Solid lines indicate the luminosity distance PDF for
unlensed sirens. Gravitational lensing introduces asymmetry and
additional dispersion to the observed luminosity distance PDF as can
be seen from the dotted curves.
If the observed distances are corrected for gravitational lensing,
both dispersion and asymmetry can be significantly reduced. Corrected
PDFs are indicated by dashed lines. 

Although the mean magnification factor is unity, the mean of $\mu^{-1/2}$
might be different from unity depending on the moments of the PDF of
$\mu$. The mean luminosity distance of a
lensed or corrected standard siren might thus be biased compared to
an unlensed siren. 
This bias can be circumvented if $D_{\rm L}^{-2}$ is
considered instead of $D_{\rm L}$ (see Section \ref{sec:flux}). 
For a lensed standard siren at $z=1$ the mean
luminosity distance is overestimated by 
$\sim 0.1\%$ and this effect
increases with redshift to $\sim 0.6\%$ at $z=4$. Correcting for gravitational
lensing reduces the bias at $z \lesssim 3$, but increases the bias
at $z=4$ where the mean of the corrected
luminosity distance is overestimated by $\sim 1\%$.
However, since we expect only one or a few standard sirens at each redshift, 
the location of the mode is more important than the mean. As can
be seen in Figure \ref{fig:pdf}, the mode of the corrected luminosity 
distance PDF is always closer to the true unlensed distance than the mode of 
the observed (lensed) luminosity distance.

After correcting for gravitational
lensing, we are left with an absolute standard siren with an almost
Gaussian dispersion of $\delta D_{\rm L}/D_{\rm L} \sim 1\%$ and
$\delta D_{\rm L}/D_{\rm L} \sim 3\%$ at redshift $z=1$ and $z=4$,
respectively.

\subsection{Distance Error from Position Uncertainty}
The accuracy to which the position of the standard siren could be
determined affects not only the distance determination, but also the
uncertainty of the estimation of the magnification factor. If the
position of the siren could be narrowed down to the host galaxy only,
the dispersion in the corrected luminosity distance 
could be significantly increased. An uncertainty of 3
arcseconds in the position of the source increases the dispersion in
the corrected luminosity distance with 10\% and 60\% 
at redshift $z=1$ and $z=4$, respectively. 
In the following, we assume that the position of the
source is pinpointed with enough precision for this source of error
not to dominate. Note, however, that this corresponds to
arcsecond resolution at $z=4$ which could be an optimistic assumption.

\subsection{Groups, Clusters, and Dark Halos \label{sec:groups}}
In the previous sections, we have assumed a one-to-one relation
between luminous galaxies and dark matter halos. If this assumption is
relaxed, our ability to
predict lensing magnifications could potentially be severely impaired. 
This could be caused by, e.g. completely dark halos whitout any galaxies, 
and dark matter halos hosting groups or clusters of galaxies. 
We have investigated the size of this effect by simulating lines-of-sight
using publicly available halo and galaxy catalogs constructed from
the results of N-body simulations \citep{kau99}.
Particles in the simulations which were not bound to any halo 
were treated as if
they belonged to a smooth matter component (see \citet{gun06} for
details). 
To compute the true magnification factors, data from the halo
catalogs, including halo masses, were used. 
When the magnification factor was estimated, data from the galaxy
catalogs were used and halo masses were obtained from galaxy luminosities 
following the prescription in \citet{gun06}. However, we used a different
Tully--Fisher relation \citep{gio97} in order to match the
normalization of the models used by \citet{kau99}. In the calculations
of the magnification factors all halos were parametrized as truncated 
NFW-profiles.

We have studied the effects of groups, clusters, and dark halos on 
our ability to predict the magnification of a source at $z=1.5$. 
For a source at this redshift the dispersion, 
in terms of the root mean square deviation, of the PDF of $\mu$ is 
$\sim 7\%$ and the corresponding dispersion in the luminosity distance
PDF is $\sim 3\%$. If the magnification could be recovered exactly, 
the dispersion in $\mu/\mu_{\rm est}$ would vanish.

The effect of dark halos is partly taken into consideration in 
\citet{gun06}
where galaxies fainter than the magnitude limit play the role of completely
dark halos. However, to make an extremely conservative estimate, we   
use simulated lines-of-sight from N-body results where only the most massive, 
but not all, of the halos contain galaxies \citep{kau99}. 
The contribution to the dispersion in $\mu/\mu_{\rm est}$ due to 
the $\sim 90\%$ of the halos in the simulated lines-of-sight which are
not associated to any galaxy and thus cannot be modelled, is $\sim 3\%$.

Approximately $2\%$ of the dark matter halos host more than one galaxy,
corresponding to groups or clusters of galaxies. Even though this
number is low, 
since these halos are also the most massive ones, the effect of poor modelling 
of groups of galaxies could potentially add a sizeable dispersion. 
Our simulations 
show that these halos add a dispersion of $\sim 3\%$ to the dispersion of 
$\mu/\mu_{\rm est}$. However, since
groups and clusters of galaxies could be identified, better modelling
of the halos of these systems might be used, which would reduce the dispersion.
For lines-of-sights in the outskirts of clusters even lensing maps 
constructed from multiple image systems might be used to estimate the 
magnification. 

If a $\sim 50\%$ scatter in the relation between galaxy luminosity and
mass, completely dark halos, and poor
modelling of galaxy groups
are taken into account, the dispersion in the magnification
factor can be reduced from $\sim 7\%$ to $\sim 4\%$. This translates into an
improvement in the dispersion in the luminosity distance from $\sim 3\%$ to
$\sim 2\%$. Such a reduction of the dispersion corresponds to increasing
the sample size by almost a factor $3$.
However, the mean of the corrected luminosity distance is underestimated by
$\sim 1\%$, considerably worse than the bias of $\sim 0.2\%$ toward 
higher mean in the case of observed (lensed) luminosity distance.

A biased relation between galaxy luminosity and halo mass 
could severely impair our capability to correct for
gravitational lensing. If all masses were 
underestimated by $\sim 50\%$, the dispersion in the 
magnification factor and luminosity distance could be 
reduced to $\sim 5\%$ and $\sim 2\%$, respectively.
Since the the mean of the corrected luminosity distance is 
underestimated by only $\sim 0.5\%$ in this case, such a bias in the
luminosity-mass relation would not significantly hamper or ability to
correct for lensing.

If, on the other hand, all masses were
overestimated by $\sim 50\%$, correcting for gravitational
lensing would instead increase the dispersion in $\mu/\mu_{\rm est}$ 
to $\sim 10\%$, ruining our attempt to correct for gravitational lensing.
Since we require the total mass in halos to be $\leq \Omega_{\rm M}$,
a too large bias toward higher halo masses would be discovered.
Furthermore, any substantial bias in the luminosity-mass relation
should give worse residuals in the Hubble diagram after correcting data for 
lensing than before.

In the so called realistic scenario outlined in \citet{gun06}, where several
effects are taken into account, the dispersion in 
$\mu/\mu_{\rm est}$ for a source at $z=1.5$ is $\sim 3\%$. 
The discussion above lead us to the conclusion that in order for
lensing corrections to be useful, the
dispersion in $\mu/\mu_{\rm est}$ must be $\lesssim 5\%$, leaving us
with an error budget of $\sim 4\%$ for dark halos and groups and 
clusters of galaxies. Since our extremely conservative analysis yields
$\lesssim 4\%$, we are confident that this number is within reach.

\section{IMPROVING COSMOLOGICAL PARAMETER ESTIMATION}

To illustrate the improvement if corrections for gravitational lensing
are applied, we have fitted cosmological parameters to simulated data
sets and plotted the corresponding confidence contours.

\subsection{Simulated Data Sets}\label{sec:data}
Our fiducial supermassive binary black hole data set consists of 10
luminosity distance measurements to standard sirens evenly distributed
in the redshift range $1 \leq z \leq 4$ with errors as described in
Section~\ref{sec:dpdf} (high redshift sample). In addition, chirping
neutron star binaries might provide gravitational wave standard sirens
with optical counterparts at low redshifts \citep{dal06}.  We assume 5
additional distance measurements at low redshifts, $0.04 \leq z \leq
0.12$ (low redshift sample). The fractional error in these distances
were assumed to be given by $\delta D_{\rm L}/D_{\rm L}=D_{\rm
L}^2/(1700 \, {\rm Gpc})$ adopted from
\citet{dal06}. Since the improvement from correcting
for lensing corresponds to a factor of three in sample size almost
independently of the exact rate and redshift distribution of standard
sirens (see Section \ref{sec:flux}), 
the fractional improvement in the confidence contours can be
appreciated although the absolute number of well observed sirens is
highly uncertain.

\subsection{Maximum Likelihood Technique}
Since the PDFs of lensed distances are expected to be asymmetric (see
Figure \ref{fig:pdf}), a maximum likelihood technique is used to fit
cosmological parameters to the simulated data. Given a data set of
redshifts and luminosity distances, $\{z_i,D_i\}$, obtained from
gravitational wave standard sirens and their electromagnetic counterparts,
cosmological parameters can be estimated by maximizing the logarithm
of the likelihood,
\begin{equation}
\log L(\theta)=\sum_i \log f(D_i;z_i,\theta),
\end{equation}
where $\theta$ represents the cosmological parameters whose values we
want to estimate.  We fit parameters to data sets consisting of
unlensed ($D=D_{\rm L}$), lensed ($D=\mu^{-1/2} D_{\rm L}$), and
corrected ($D=(\mu / \mu_{\rm est})^{-1/2}D_{\rm L}$) luminosity
distances.
For each data set the corresponding PDF, $f(D;z,\theta)$, is used.
Since the effects of gravitational lensing are redshift dependent,
PDFs for lensed and unlensed luminosity distances are redshift
dependent too.  Low redshift sirens, with negligible magnification due
to lensing, are assumed to be described by a Gaussian PDF.

In the cosmology fits, a model for luminosity distances in a flat
universe is used,
\begin{equation}
D_{\rm L}(z;h,\Omega_{\rm M},w_0,w_{\rm a})=(1+z)
\int_0^z\frac{dz'}{H(z')}.
\end{equation}
This model depends on the Hubble constant, 
$H_0=100h$ km s$^{-1}$ Mpc$^{-1}$, the matter density,
$\Omega_{\rm M}$, and the equation of state parameter of dark energy
parametrized by \citep[e.g.][]{lin03} 
$w(z)=w_0+w_{\rm a}z/(1+z)$, via the Hubble parameter,
\begin{equation}
H^2(z)=H_0^2 \left[\Omega_{\rm M}(1+z)^3+(1-\Omega_{\rm
    M})(1+z)^{3(1+w_0+w_{\rm a})}e^{-3w_{\rm a}z/(1+z)}\right]. 
\end{equation}
Note that chirping binaries, in 
contrast to e.g.~Type Ia supernovae, are absolute distance indicators, 
i.e.~we don't have to marginalize over any non-cosmological parameters. 
We have considered the effects on confidence level contours for both
constant and time varying dark energy equation of state parameters. 
In the cases where the equation of state parameter was assumed to be constant, 
the parameter $w_{\rm a}$ was put to zero and 
one of the three remaining parameters was either marginalized over or assumed 
to be exactly known.
When a time varying equation of state parameter was considered, prior
information on $\Omega_{\rm M}$ and $h$ was used.

\subsection{Effects on Cosmology Fits}
Figure \ref{fig:fits} shows the results of the cosmology fits where
a constant equation of state parameter was assumed, i.e.~$w_{\rm a}=0$. 
Each panel in the figure shows 68.3\% confidence level contours ($\log
L(\theta)=\log L_{\rm max}-2.30/2$) for unlensed (solid contours),
lensed (dotted contours), and corrected (dashed contours) data sets.
Each column of panels in the figure shows the results of cosmological
parameters fitted to different data sets. The left column shows
results obtained for the high redshift sample. The middle column
shows the effect of adding the low redshift sample. The right column
of panels shows the results of cosmology fits where the third
parameter was assumed to be exactly known. This gives an indication of
the dependence of the size of the contours on the marginalization over
and prior knowledge of the third parameter.

Panel a shows confidence level contours in the $[\Omega_{\rm
M},w_0]$-plane obtained for a simulated data set consisting of 10 high
redshift standard sirens. Gravitational lensing degrades the data set
and the cosmological results that can be obtained from it. However, the size
of the confidence level contours can be significantly reduced if the
data are corrected for lensing magnification. 
Distance measurements at low redshifts are much more useful 
in constraining $h$ than
distance measurements at high redshifts. Adding low redshift distance
measurements to the high redshift data set thus have effects similar
to applying prior information on $h$. Panel b shows the improvement
on the confidence level contours due to 5 additional distance
measurements at low redshifts. Panel c shows the very tight
constraints in the $[\Omega_{\rm M},w_0]$-plane which could be
obtained if $h$ was exactly known.

The middle row of panels shows results similar to those presented in
the top row of panels, but in the $[h,\Omega_{\rm M}]$-plane. Panel
d shows the contours obtainable from 10 high redshift standard
sirens. The improvement due to 5 additional low redshift standard
sirens can be seen in panel e. A data set consisting of both high and
low redshift distances could potentially give constraints on
$\Omega_{\rm M}$ and $h$ on the 10\% level.  Panel f shows the
results for the case where $w_0$ is assumed to be exactly known.  Note
that assuming that the dark energy is accurately described by a
cosmological constant, as current data favors 
\citep[e.g.][]{han04},
standard sirens have the potential to give very strong constraints on
the Hubble parameter, $h$, if we correct for gravitational lensing.

The bottom row of panels show results in the $[h,w_0]$-plane.
Panel g shows confidence level contours obtained from 10
high redshift sirens. The improvement due to 5 
additional low redshift standard sirens can be seen in panel h.
Panel i shows the results for the case
where $\Omega_{\rm M}$ is assumed to be exactly known.

Results obtained from the high redshift data sets of lensed or
corrected sirens suffer from severe degeneracies.
The (one-sided) limits on $\Omega_{\rm M}$, $w_0$ and $h$ 
which could be obtained from the high redshift sample alone, 
without any priors or low redshift distance
measurements, would not be competitive compared to other 
measurements of these parameters. Priors or low redshift data seem to
be necessary for a small set of high redshift standard sirens to be
able to give competitive constraints on $\Omega_{\rm M}$, $w_0$ and
$h$. 

Figure \ref{fig:fits2} shows the results in the $[w_0,w_{\rm a}]$-plane.
Solid, dotted, and dashed lines indicate the 68.3\% confidence level
contours obtained for unlensed, lensed, and corrected data sets. 
Gaussian priors, assuming $h$ and $\Omega_{\rm M}$ to be known with 
3\% accuracy, were used in the fitting procedure. Panel a shows the
contours obtained from the high redshift data set. 
The effect of adding the low redshift data set is small and 
is shown in panel b.  
Since strong priors have already been applied, there is no dramatic 
improvement due to broken degeneracies.  
Panel c shows the contours obtained with both $h$ and $\Omega_{\rm M}$ 
exactly known. The set of 10 high redshift sirens 
is evidently not sufficient to yield very strong constraints on 
$w_0$ and $w_{\rm a}$ even in 
the case where the Hubble constant and the matter density are 
exactly known. Larger data sets than the ones considered here are 
thus needed for standard sirens to
be able to provide very strong constraints on the evolution of dark energy.

In Table~\ref{tab:cl} and~\ref{tab:cl2} 
the numerical results are summarized. The
68.3\% confidence level intervals in the tables were
obtained by finding the extremal values of the cosmological 
parameter $\theta$ 
(where $\theta$ is $h$, $\Omega_{\rm M}$, $w_0$, or $w_{\rm a}$)
fulfilling the condition $\log L(\theta)=\log L_{\rm max}-2.30/2$.

\section{FLUX AVERAGING}\label{sec:flux}
For a large number of sources, all at the same redshift, the mean
magnification is expected to be unity due to flux conservation. 
Since gravitational wave flux is inversely proportional to the luminosity
distance squared, the mean of 
$D_{\rm L,\,obs}^{-2}=\mu D_{\rm L}^{-2}$ equals $D_{\rm L}^{-2}$.
The mean of the observed luminosity distance 
is, on the other hand, expected to be biased by $0.1\%-0.6\%$
as noted in Section \ref{sec:dpdf}. This bias is similar to the one which 
arises if magnitudes are used instead of fluxes. However, whether this
bias is important or can be ignored depends on the sample size and the errors
in each distance measurement. 
In the scenario studied in this paper, where the standard sirens are
sparsely distributed in redshift, the position of the mode is more
important than the mean and we can use
$D_{\rm L}^{\rm obs}$ instead of $D_{\rm L,\,obs}^{-2}$ as our observable. 

Degradation by gravitational lensing can thus be cured by studying the 
mean of the 
luminosity distance (or inverse of the luminosity distance squared if
it is important to avoid a biased mean)
$\bar D_{\rm L}$ with uncertainty $\sigma_{\bar D_{\rm L}}$,
if the number of sources is large enough \citep{hol05b}. 
We now estimate how large this number of supermassive binary black holes is.

\begin{figure}
\begin{center}
\resizebox{1.\textwidth}{!}{\includegraphics{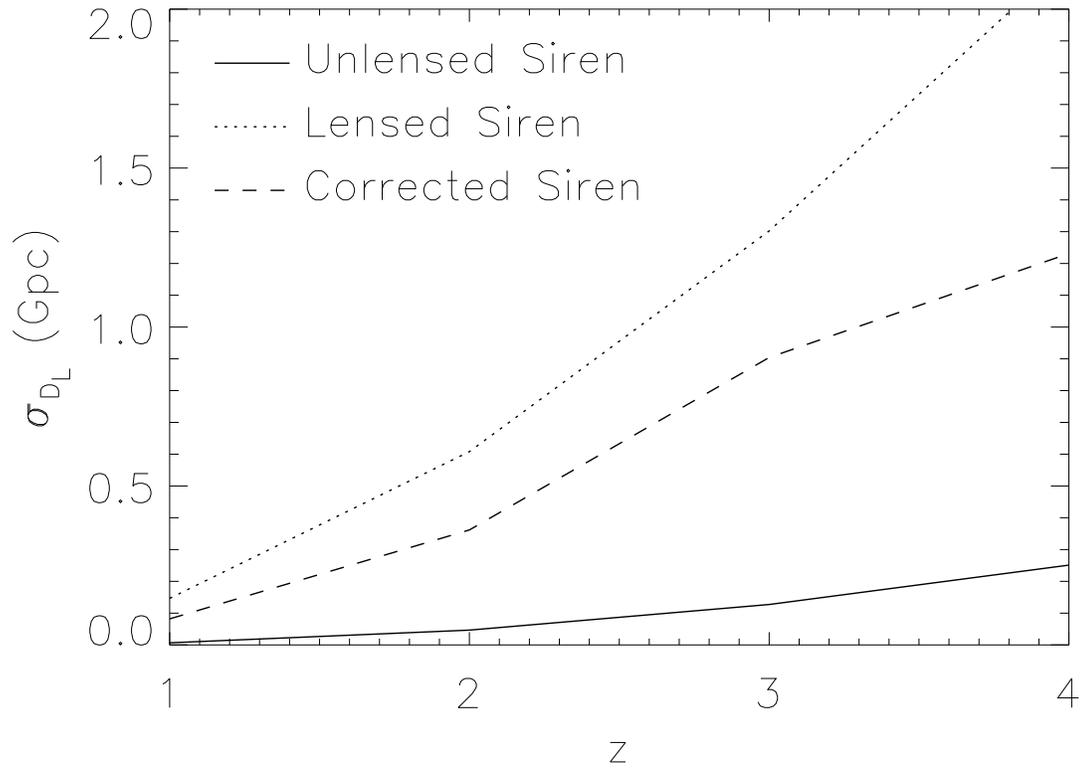}}
\caption{\label{fig:rms} Root mean square deviation of the probability
  distribution functions of unlensed, lensed, and corrected standard
  sirens as a function of redshift.}
\end{center}
\end{figure}

Figure \ref{fig:rms} shows the dispersion of the luminosity distance
PDFs as a function of redshift, quantified in terms of the root mean 
square of the deviation from the mean, $\sigma_{D_{\rm L}}$. 
The increase in the dispersion,
due to lensing, depends on the redshift. At $z=1$ and $z=4$ the
dispersion is increased a factor $\sim 20$ and $\sim 10$,
respectively. If the distances obtained from the sirens are corrected,
the dispersion can be reduced by a factor $\sim 2$.  If a number of
lensed sirens, all at the same redshift, would be observed, how many
would be needed to compensate for the dispersion due to lensing? The
dispersion of the mean value of the distances inferred from the
standard sirens should be inversely proportional to the square root of
the number of sirens,
\begin{equation}
\sigma_{\bar D_{\rm L}}=
\sigma_{D_{\rm L}}^{\rm lensed}N^{-1/2}
\end{equation}
To restore the dispersion to the value corresponding to one unlensed
siren, i.e. 
$\sigma_{\bar D_{\rm L}}=\sigma_{D_{\rm L}}^{\rm unlensed}$,
the needed number of observed sirens would thus be,
\begin{equation}
N=(\sigma_{D_{\rm L}}^{\rm
lensed}/\sigma_{D_{\rm L}}^{\rm unlensed})^2.
\end{equation} 
Since the dispersion of an unlensed siren increases much faster as a
function of redshift than the dispersion of a lensed or corrected
siren, the number of sirens needed to compensate for lensing is much
larger at $z=1$ ($N \sim 500$) than $z=4$ ($N \sim 70$). Only in very
optimistic scenarios would we be able to observe such a large number
of sources including optical counterparts.

The number of lensed sirens needed to achieve a dispersion comparable
to the dispersion of one corrected siren is 
\begin{equation}
  N = (\sigma_{D_{\rm L}}^{\rm lensed}/\sigma_{D_{\rm L}}^{\rm
  corrected})^2\sim 3 ,
\end{equation}
almost independent of redshift. Thus, unless the number of
well-observed sirens is very large, by correcting for the effects of
gravitational lensing, we are able to improve our data by an amount
that is comparable to increasing the sample size by a factor of three.

\section{SUMMARY AND DISCUSSION}
Gravitational waves emitted by chirping supermassive black hole
binaries measured by LISA could in principle provide very accurate
distance determinations. If electromagnetic counterparts of these
events could be found, from which redshifts could be determined, these
standard sirens could be used to build a high redshift Hubble
diagram. In principle, chirping binaries are almost perfect distance
indicators, only limited by the precision of the position
measurement. Thus, the need of controlling systematic errors on the
distance measure is crucial to be able to use the standard sirens to
their full potential. These errors will most likely be completely
dominated by gravitational lensing.

\begin{figure}
\begin{center}
\resizebox{0.32\textwidth}{!}{\includegraphics{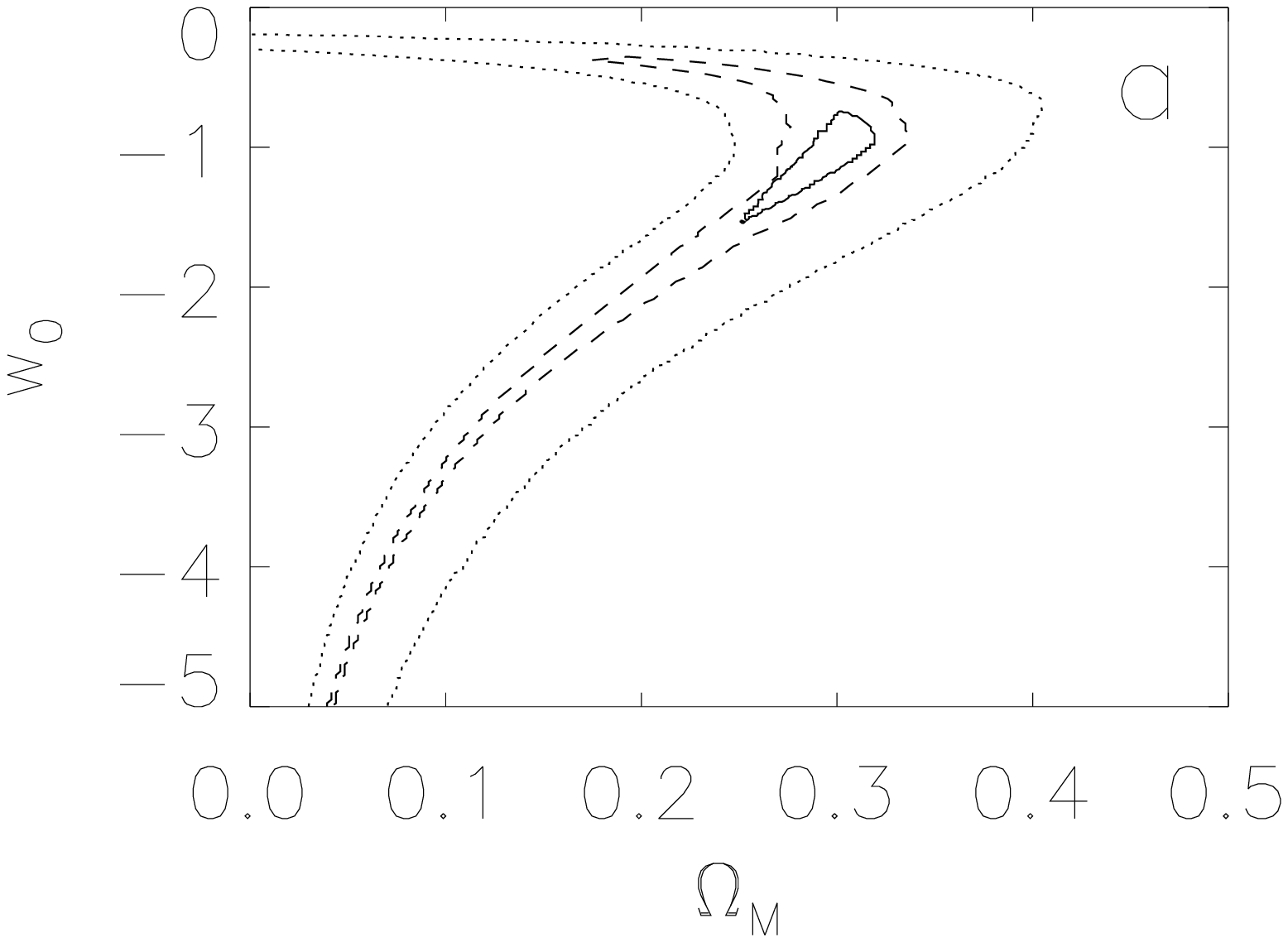}}
\resizebox{0.32\textwidth}{!}{\includegraphics{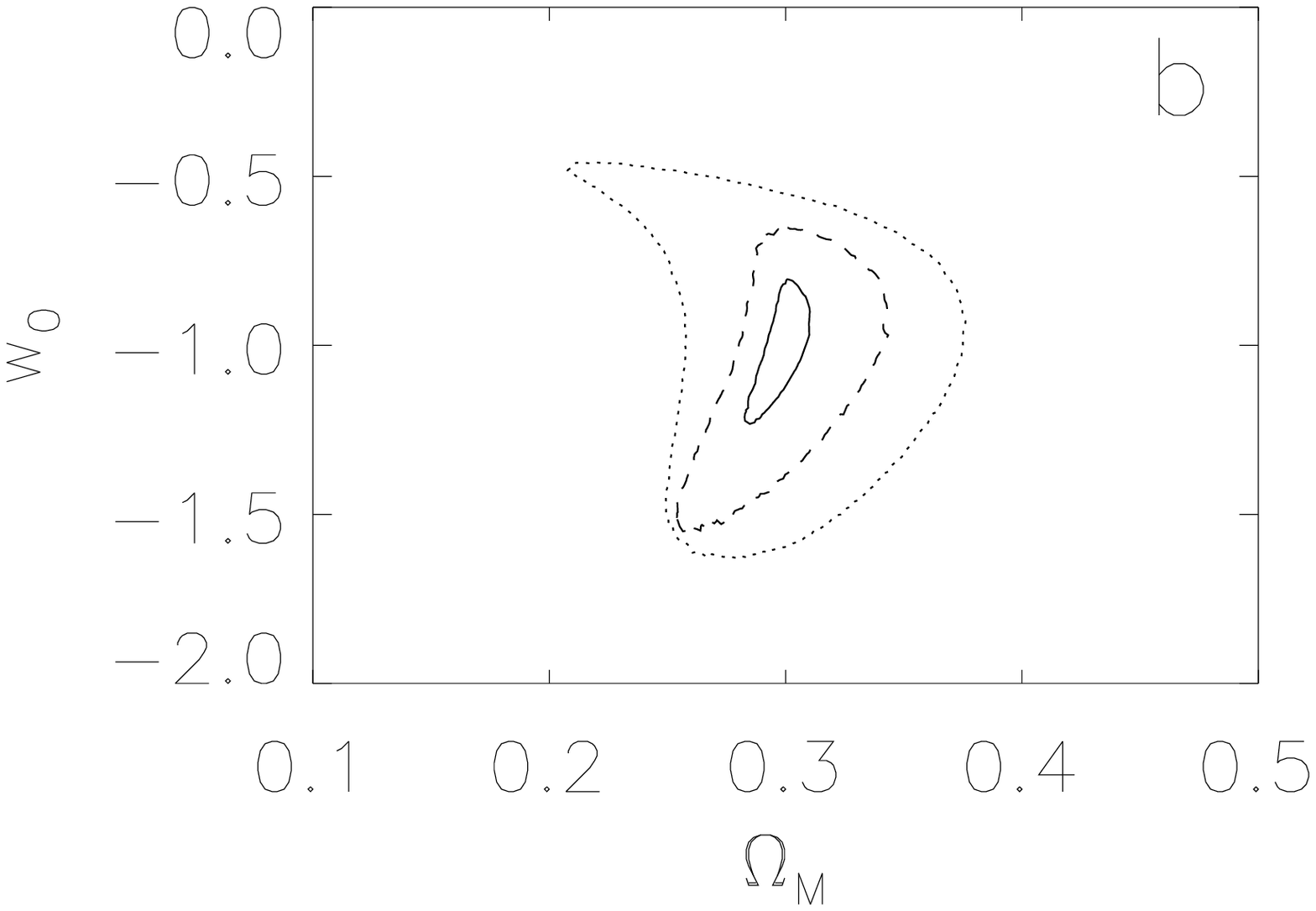}}
\resizebox{0.32\textwidth}{!}{\includegraphics{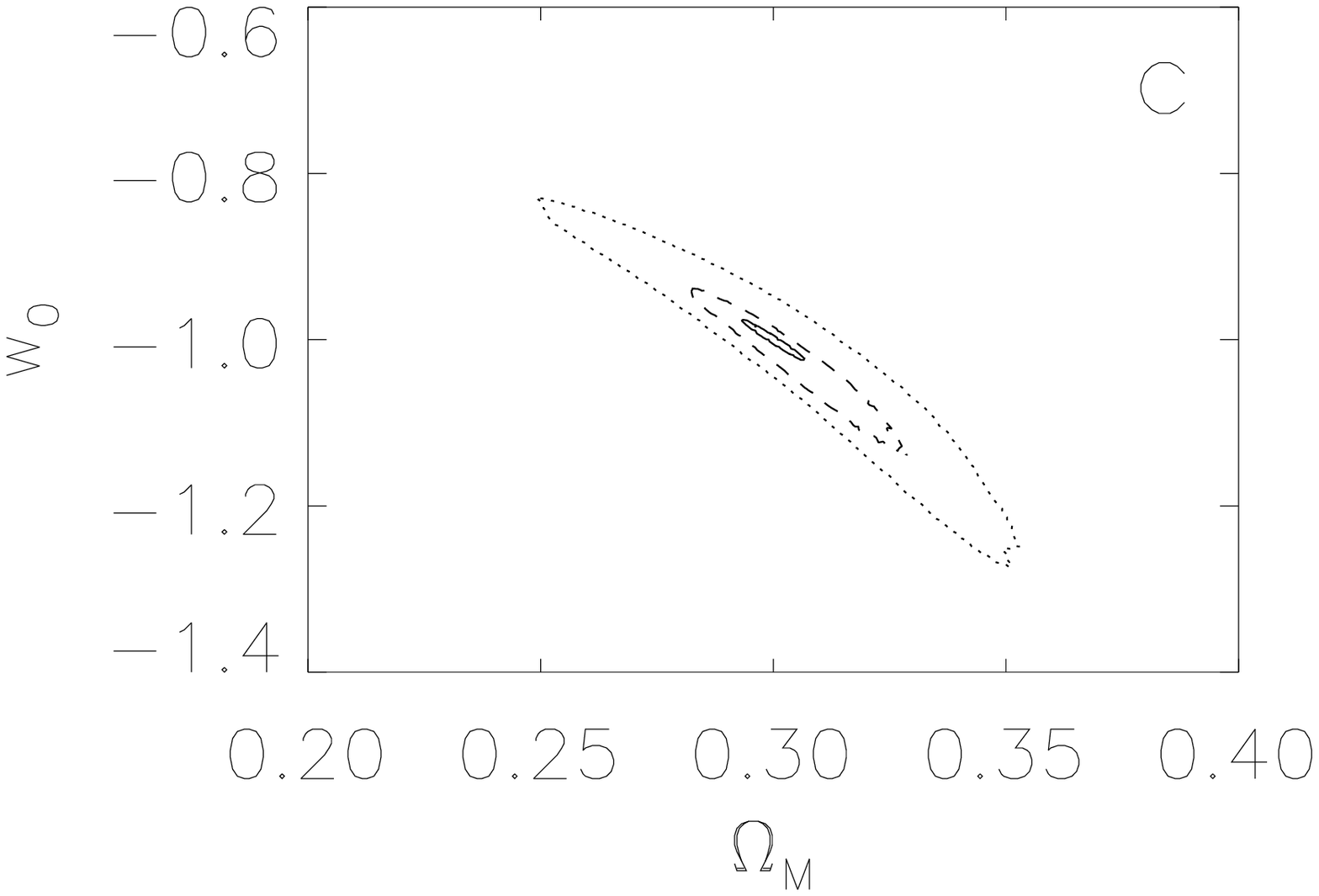}}
\resizebox{0.32\textwidth}{!}{\includegraphics{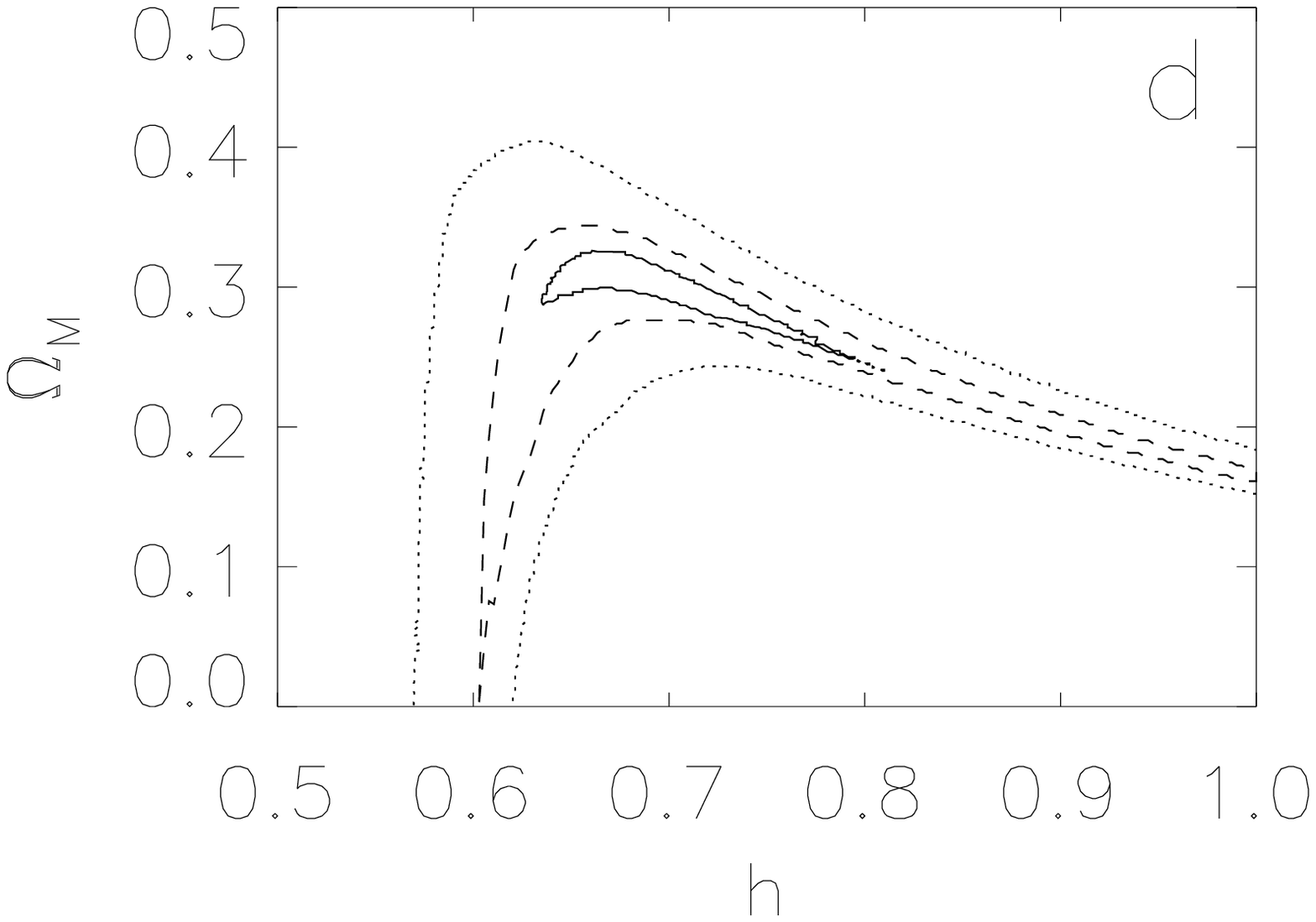}}
\resizebox{0.32\textwidth}{!}{\includegraphics{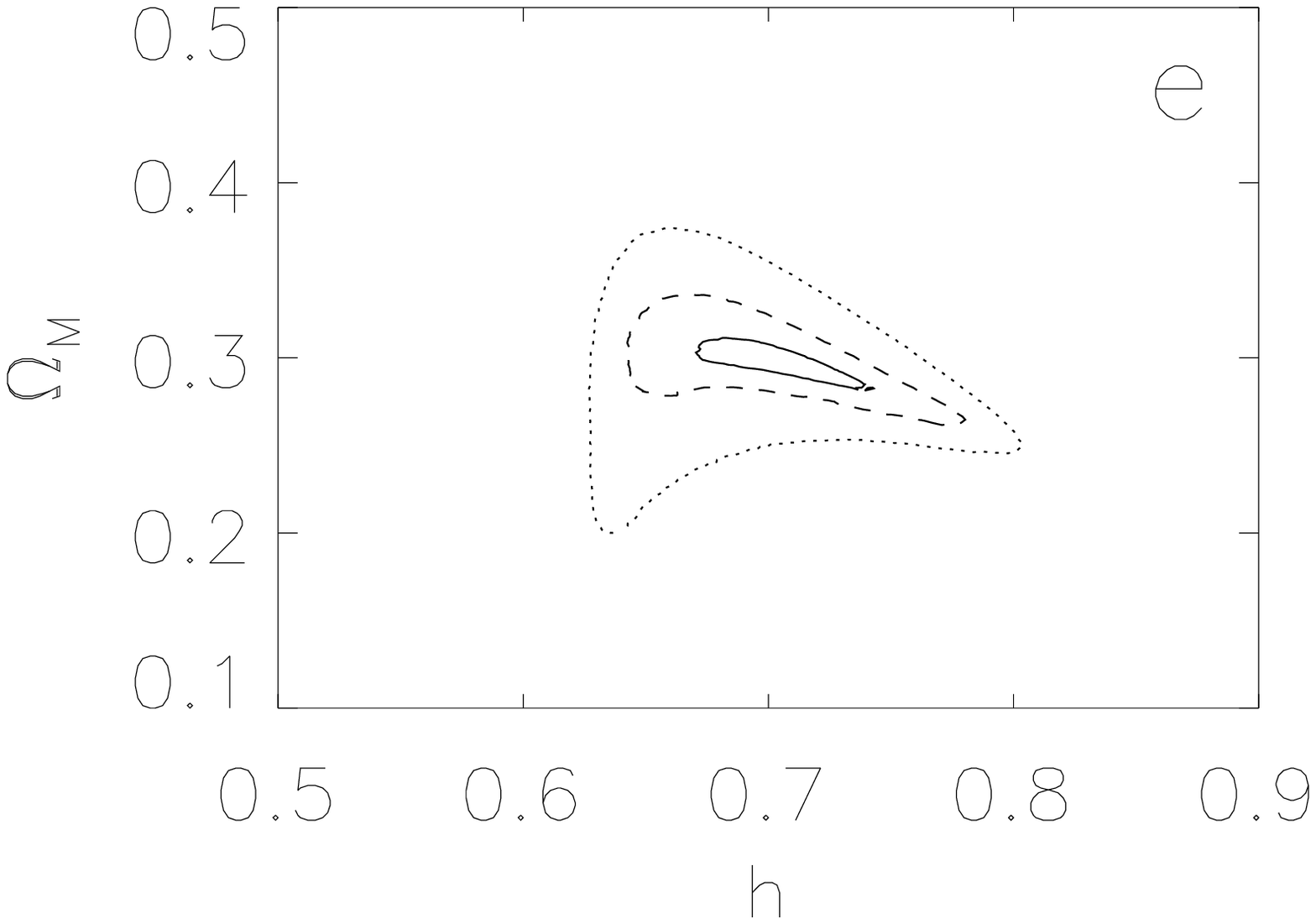}}
\resizebox{0.32\textwidth}{!}{\includegraphics{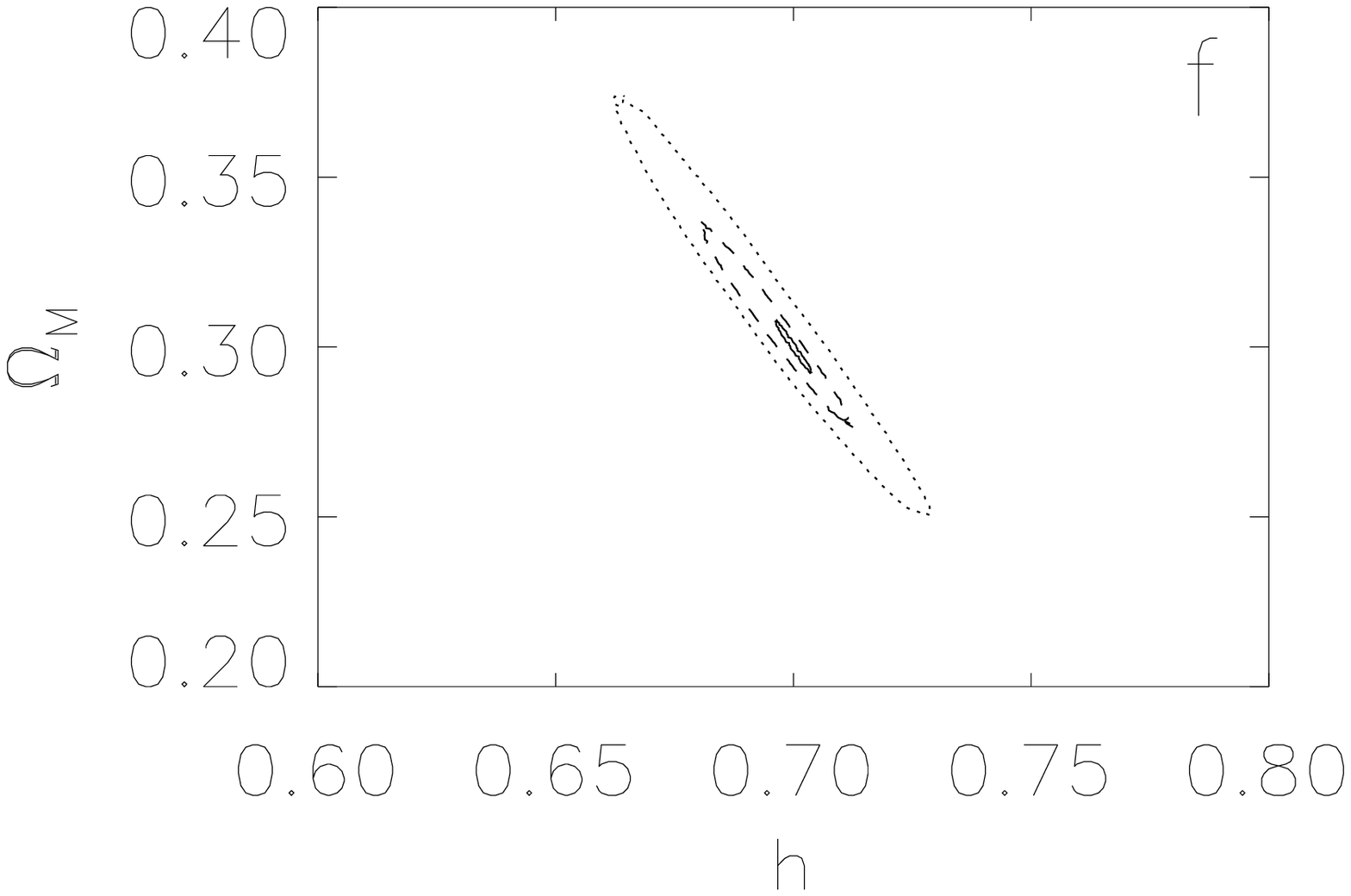}}
\resizebox{0.32\textwidth}{!}{\includegraphics{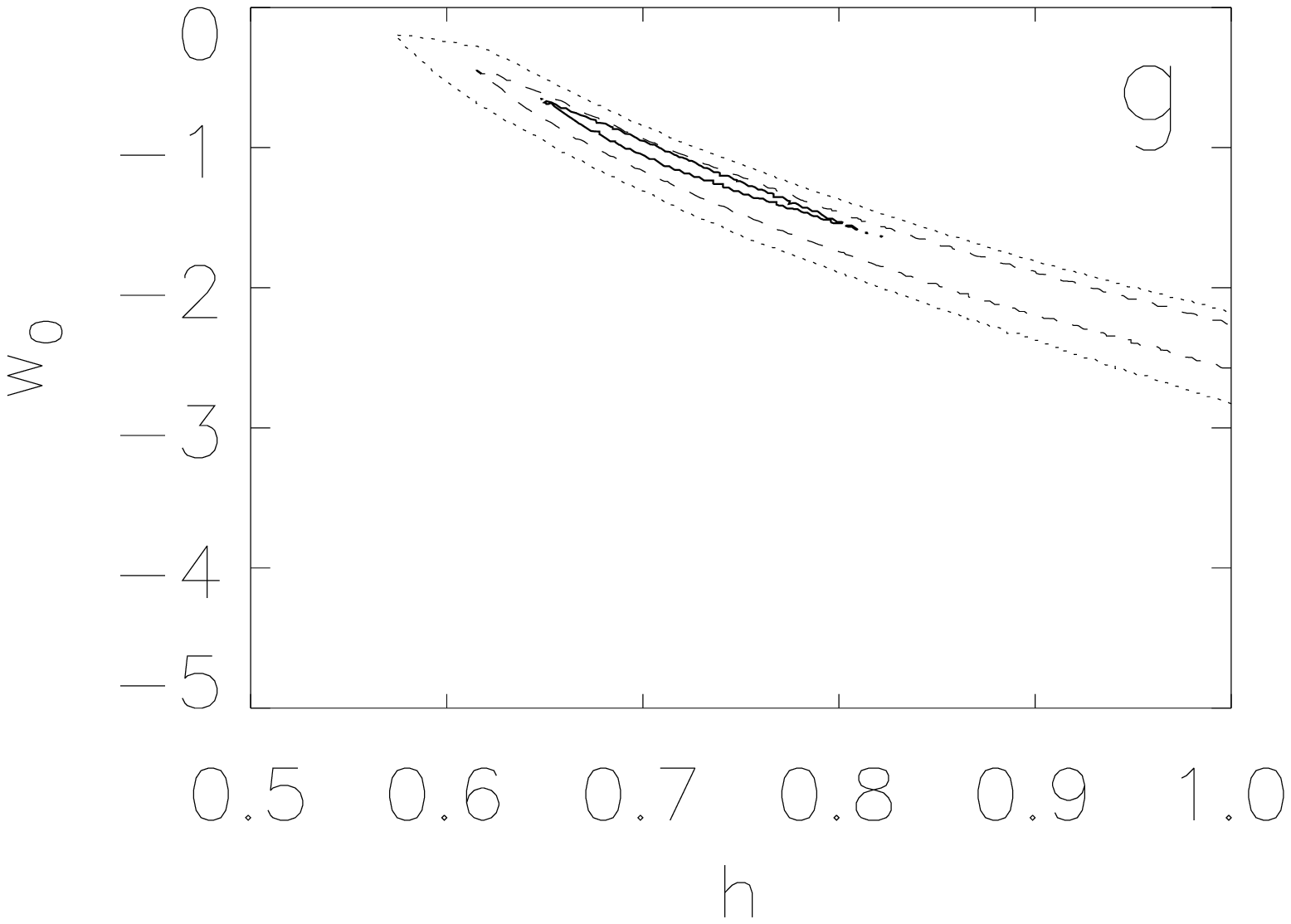}}
\resizebox{0.32\textwidth}{!}{\includegraphics{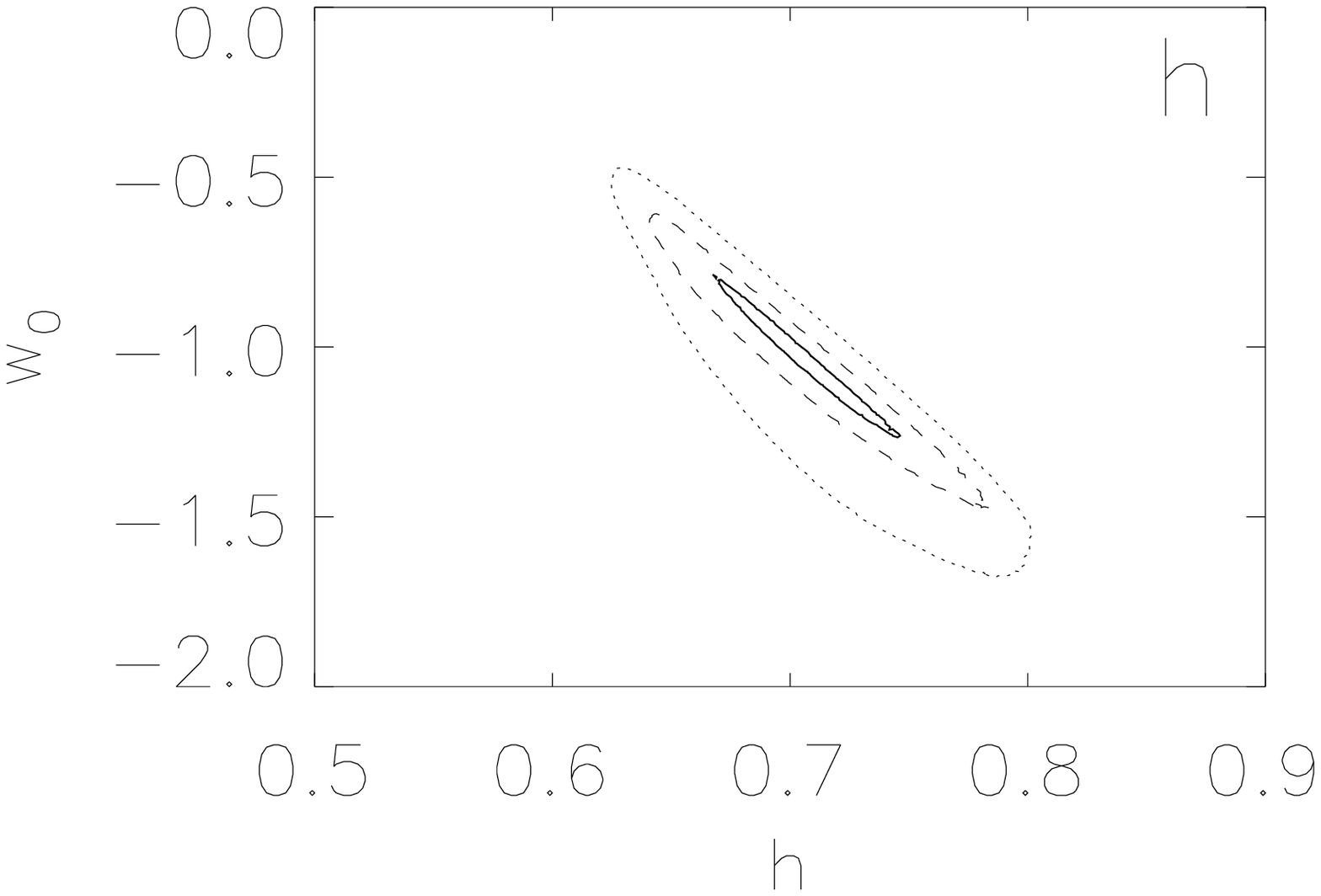}}
\resizebox{0.32\textwidth}{!}{\includegraphics{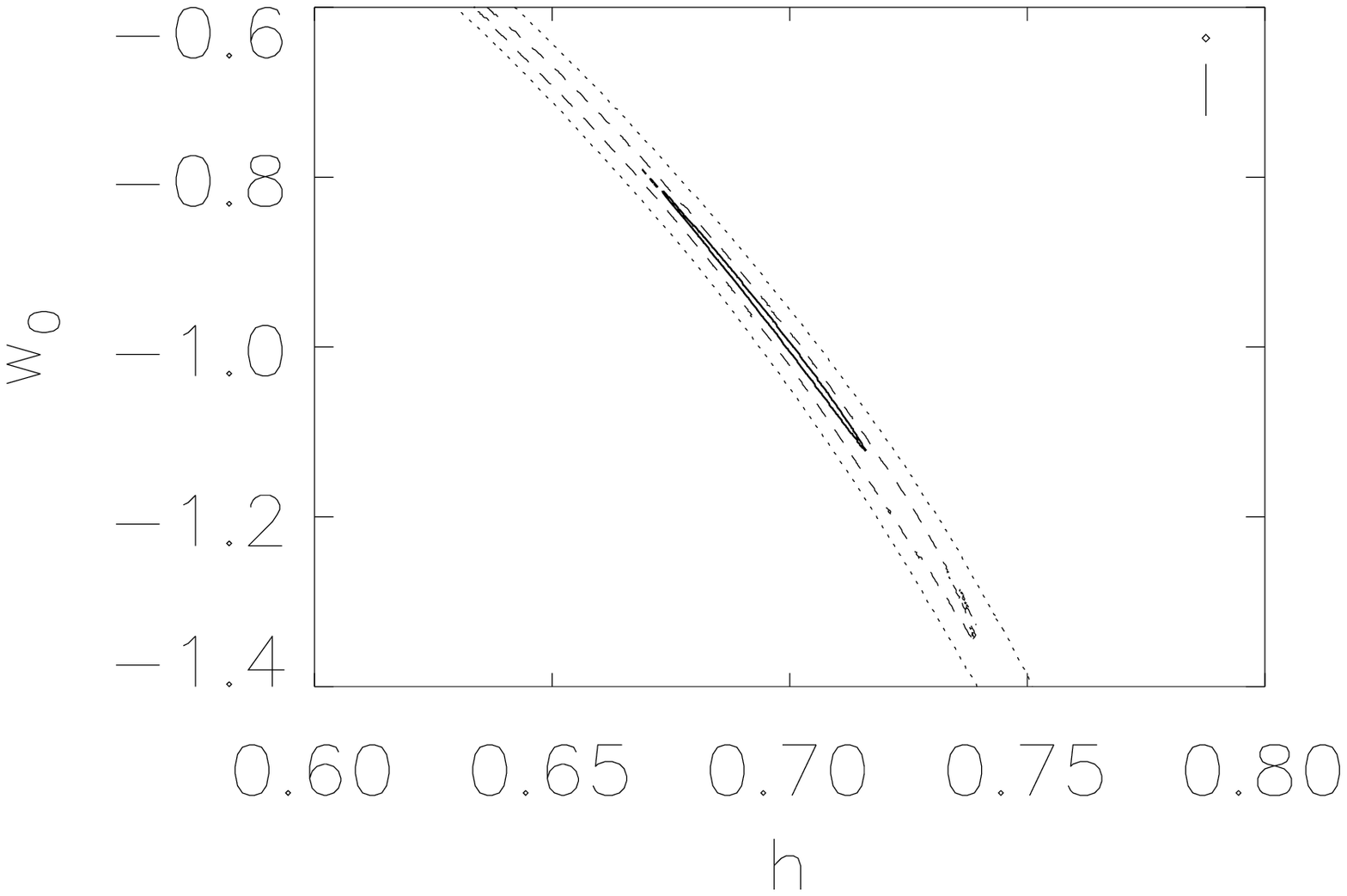}}
\caption{\label{fig:fits} Confidence level contours at the 68.3\%
  confidence level for different cosmological parameters and different
  simulated data sets. 
  Solid, dotted, and dashed contours corresponds to unlensed, lensed,
  and corrected data sets, respectively.
  Panel a: High redshift data set. Panel b: High and low redshift data
  sets. Panel c: High redshift data set but $h$ is known exactly.
  Panel d: High redshift data set. Panel e: High and low redshift data
  sets. Panel f: High redshift data set but $w_0$ is known exactly.
  Panel g: High redshift data set. Panel h: High and low redshift data
  sets. Panel i: High redshift data set but $\Omega_{\rm M}$ is
  known exactly.
  Note the different scales in the figures.
}
\end{center}
\end{figure}

\begin{figure}
\begin{center}
\resizebox{0.32\textwidth}{!}{\includegraphics{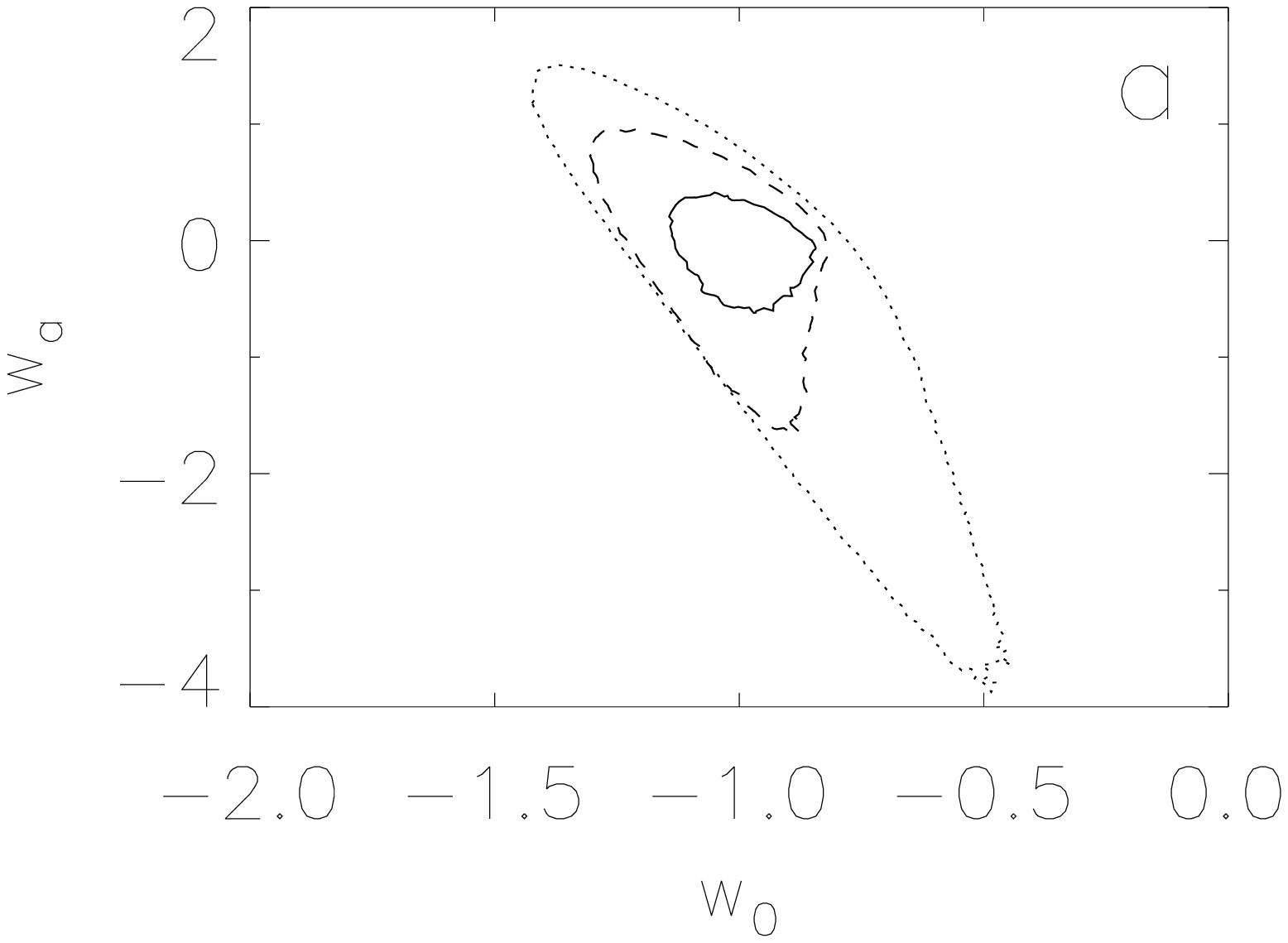}}
\resizebox{0.32\textwidth}{!}{\includegraphics{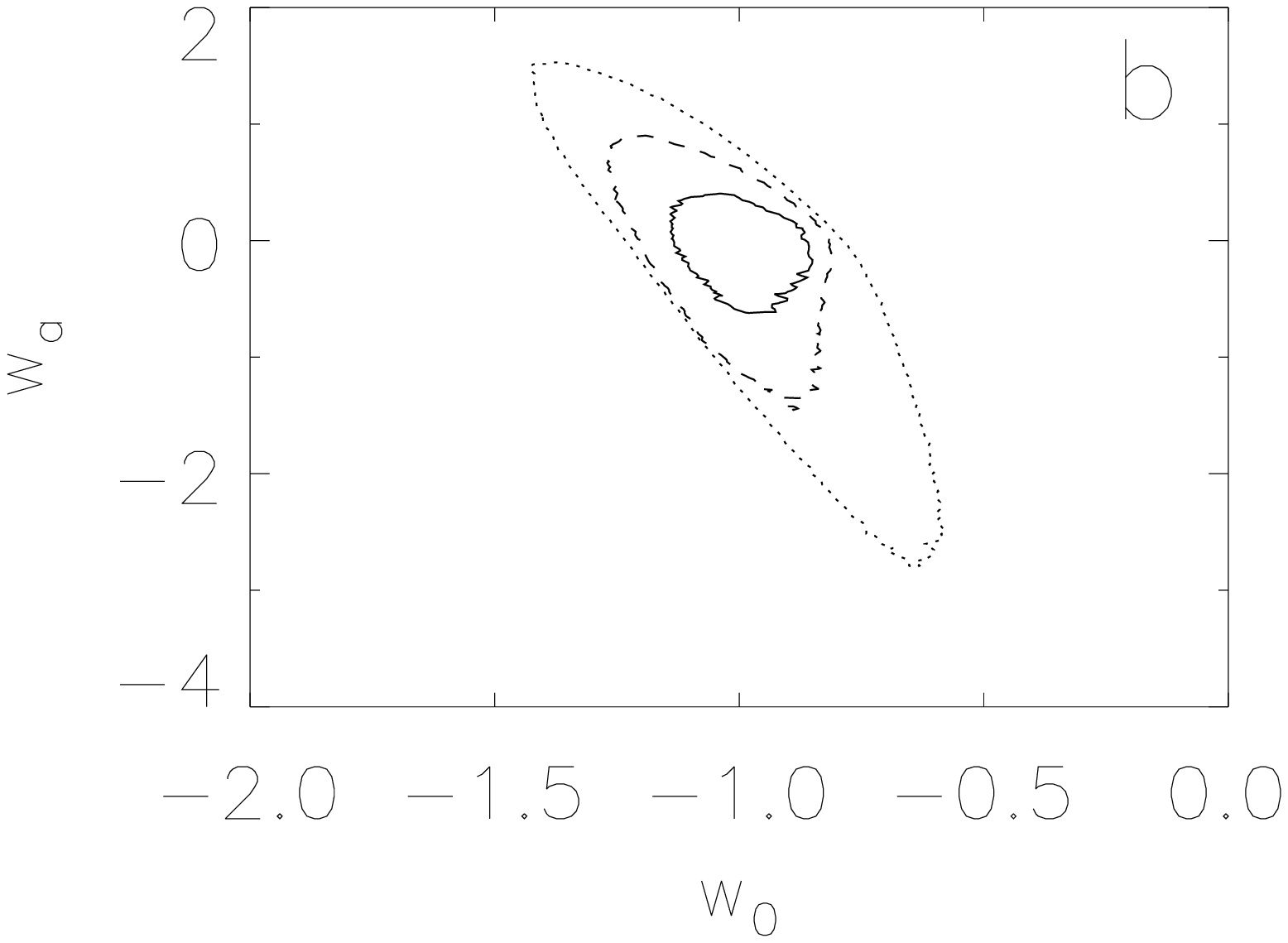}}
\resizebox{0.32\textwidth}{!}{\includegraphics{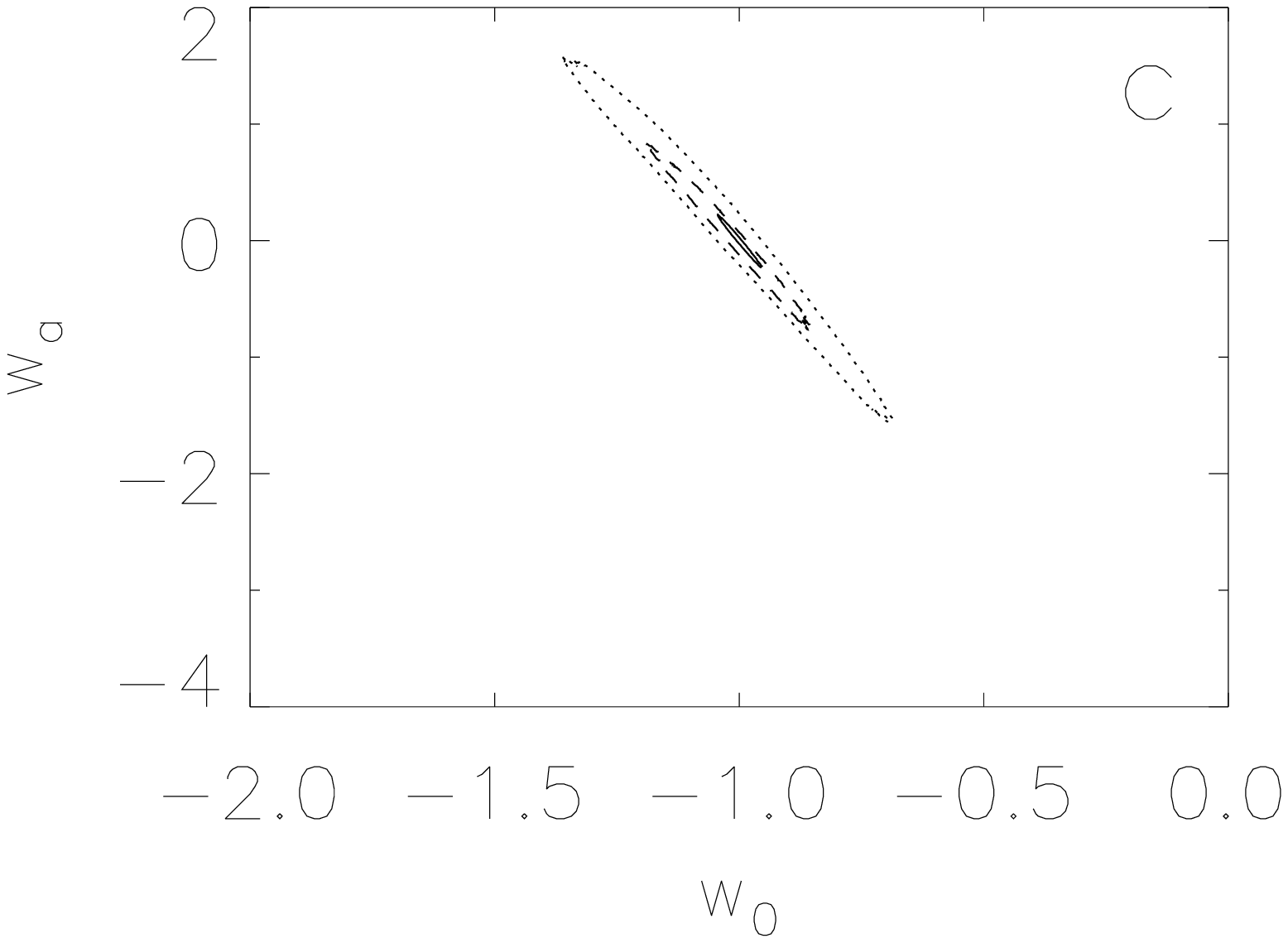}}
\caption{\label{fig:fits2} Confidence level contours at the 68.3\%
  confidence level in the $[w_0,w_{\rm a}]$-plane for different data sets. 
  Solid, dotted, and dashed contours corresponds to unlensed, lensed,
  and corrected data sets, respectively. 
  Panel a: High redshift data set. Panel b: High and low redshift data
  sets. Panel c: High redshift data set but $h$ and $\Omega_{\rm M}$
  are known exactly.
  Gaussian priors (3\% accuracy) on $h$ and $\Omega_{\rm M}$ 
  were used in the fitting procedure. }
\end{center}
\end{figure}

Since the number of expected events is uncertain and whether optical
counterparts exist or can be observed is even more unclear, we have
reasons to believe that the number of events which could be put in a
Hubble diagram is low. Thus, we can not be certain that the degradation due
to gravitational lensing can be overcome by large statistics.  From
Figure \ref{fig:fits} and \ref{fig:fits2}, it is obvious that 
correcting for gravitational
lensing can be crucial for the competitive use of gravitational wave
standard sirens in cosmology. Note also that the actual number of
observed sirens with electromagnetic counterparts at very high redshift
($z>4$) might be substantial \citep{dot06}, in which case the effects
from gravitational lensing will be even larger.

Low redshift points in the Hubble diagram (where the luminosity
distance mainly is sensitive to the Hubble constant) are important to
break some of the degeneracies on the cosmological parameters
occurring for high redshift luminosity distances, but no supermassive
black hole binaries are expected at low redshift. The situation is
thus very similar to the case of using gamma ray bursts as standard
candles where, apart from the difficulty of calibrating the candles,
the lack of low redshift sources weakens their power to constrain
cosmological parameters, see \citet{mor05}. Since chirping
supermassive black hole binaries are absolute standard candles
however, their usefulness for cosmology could be increased if they are
combined with other absolute distance determinations in the local
universe.

Besides constraining the distance--redshift relation, standard sirens
have other potential applications in cosmology.
Since the dispersion from gravitational lensing depends on the matter
distribution in the universe, it could in principle be used to
constrain cosmological parameters. Equivalent to the use of galaxy
shear measurements, the lensing magnification PDF can provide
information on $\sigma_8$ and $\Omega_{\rm M}$. The feasibility of the
method has been studied in the context of future Type Ia supernova
observations in \citet{coo06} and \citet{dod05}. In this paper, we
only note that standard sirens have the potential to be even better
probes of the cosmic magnification, due to the much smaller (at least
a factor of ten) intrinsic scatter in the luminosity
\citep{coo06}. Also, the large distances to the standard siren
sources makes it possible to study the matter distribution in the
universe at high redshifts.  However, whether this potential can be
realized depends upon the very uncertain detectability rate and the
possibility to secure redshifts for the sources. 

Even if the number of observed sirens turns out to be low, the small
intrinsic dispersion combined with the fact that chirping binaries are
absolute standard candles allows us to measure the magnification for
individual sources and thus probe the matter content, i.e., the
convergence $\kappa$ along individual lines-of-sight. This could yield
important information on the matter distribution in clusters and
galaxies. Specifically, it could be used to break the mass sheet
degeneracy that plagues galaxy shear investigations.

In conclusion, we note that although detection rates are very
uncertain, LISA could in principle provide high precision distance
measurements to very distant objects in the Universe, the error of which
will be dominated by gravitational lensing. Correcting for these
errors is comparable to increasing the sample size by a factor of
three which decreases the error contours for cosmological parameters
substantially. Depending on the exact redshift distribution of the
sirens, this could yield competitive constraints on, in particular, the
Hubble constant and the matter density, but perhaps also on dark
energy properties.

\acknowledgments
The authors would like to thank Tomas Dahl\'en and Christofer
Gunnarsson for help with the gravitational lensing simulations and an
anonymous referee for constructive criticism. 
AG would like to acknowledge support by the Swedish Research Council and 
the G\"oran Gustafsson Foundation for
Research in Natural Sciences and Medicine.

\clearpage

\clearpage

\clearpage

\clearpage

\begin{deluxetable}{llccc} 
\tabletypesize{\scriptsize}
\tablecaption{A summary of the results of the cosmology fits 
 shown in Figure~\ref{fig:fits}.
 Quoted 68.3\% confidence level intervals for 
 $\Omega_{\rm M}$, $w_0$, and $h$ were obtained by finding the
  extremal value of the cosmological parameter $\theta$ where 
  $\log L(\theta)=\log L_{\rm max}-2.30/2$. \label{tab:cl}}
\tablewidth{0pt}
\tablehead{
\multicolumn{2}{c}{Simulated Data Set}  & 
\colhead{$\Omega_{\rm M}$} & \colhead{$w_0$} & \colhead{$h$}
} 
\startdata
High-z 
& Unlensed   &
$[0.25,0.32]$ & $[-1.6,-0.7]$ & $[0.64,0.82]$  \\
& Lensed     & 
$<0.40$ & $<-0.2$ & $>0.57$  \\
& Corrected  & 
$<0.34$ & $<-0.4$ & $>0.61$  \\
High-z + low-z  & 
Unlensed &
$[0.28,0.31]$ & $[-1.2,-0.8]$ & $[0.67.0.75]$ \\
 & Lensed &  
$[0.20,0.38]$ & $[-1.6,-0.5]$ & $[0.63.0.80]$ \\
& Corrected &  
$[0.26,0.34]$ & $[-1.5,-0.6]$ & $[0.64.0.78]$ \\
High-z  + $h$  known & Unlensed &
$[0.29,0.31]$ &  $[-1.02,-0.98]$ &  \nodata \\
 & Lensed  &
$[0.25,0.35]$ & $[-1.27, -0.83]$ &  \nodata \\
& Corrected & 
$[0.28,0.33]$ & $[-1.14, -0.94]$ & \nodata \\
High-z  + $\Omega_{\rm M}$  known & 
Unlensed   & 
 \nodata & $[-1.12, -0.79]$ & $[0.67,0.72]$  \\
 & Lensed  & 
 \nodata &  $[-1.67,-0.38]$ & $[0.59,0.77]$  \\
& Corrected   &  
 \nodata & $[-1.34,-0.56]$ & $[0.63,0.74]$  \\
High-z  + $w_0$  known & 
Unlensed &
$[0.29,0.31]$ & \nodata & $[0.696,0.704]$ \\
 & Lensed   & 
$[0.25,0.37]$ & \nodata &  $[0.66,0.73]$  \\
& Corrected &
$[0.28,0.34]$ & \nodata & $[0.68,0.71]$  
\enddata

\end{deluxetable}

\clearpage

\begin{deluxetable}{llcc} 
\tabletypesize{\scriptsize}
\tablecaption{A summary of the results of the cosmology fits 
  shown in Figure~\ref{fig:fits2}.
  Quoted 68.3\% confidence level intervals for 
 $w_0$ and $w_{\rm a}$ were obtained by finding the
  extremal value of the cosmological parameter $\theta$ where 
  $\log L(\theta)=\log L_{\rm max}-2.30/2$. \label{tab:cl2}}
\tablewidth{0pt}
\tablehead{
\multicolumn{2}{c}{Simulated Data Set}  & 
\colhead{$w_0$} & \colhead{$w_{\rm a}$} 
} 
\startdata
High-z 
& Unlensed   &
 $[-1.13,-0.85]$ & $[-0.6,0.4]$  \\ 
& Lensed     & 
 $[-1.41,-0.44]$ & $[-3.8,1.5]$  \\
& Corrected  & 
 $[-1.29,-0.83]$ &  $[-1.6,0.9]$  \\
High-z + low-z  & 
Unlensed &
$[-1.13,-0.86]$ & $[-0.6,0.4]$  \\
 & Lensed &  
$[-1.40,-0.59]$ & $[-2.8,1.5]$  \\
& Corrected &  
$[-1.25,-0.83]$ & $[-1.4,0.9]$  \\
High-z  + $h$ and $\Omega_{\rm M}$ known & 
Unlensed &
 $[-1.04,-0.96]$   & $[-0.2,0.2]$  \\
 & Lensed   & 
 $[-1.36,-0.69]$   & $[-1.6,1.6]$  \\
& Corrected &
 $[-1.19,-0.86]$   & $[-0.8,0.8]$ 
\enddata

\tablecomments{In the fitting procedure Gaussian priors (3\% accuracy)
  on $h$ and $\Omega_{\rm M}$ were used.} 

\end{deluxetable}

\end{document}